\documentclass[12pt]{article}

\usepackage{graphicx}

\title{Nonperturbative parton distributions and the proton spin problem }
\author{  Yu.A.Simonov \\
State Research
Center\\Institute of Theoretical and Experimental Physics, \\
Moscow, 117218 Russia}

\newcommand{\beq}{\begin{eqnarray}}
 \newcommand{\eeq}{\end{eqnarray}}
\newcommand{\be}{\begin{equation}}
 \newcommand{\ee}{\end{equation}}

\def\fun#1#2{\lower3.6pt\vbox{\baselineskip0pt\lineskip.9pt
\ialign{$\mathsurround=0pt#1\hfil ##\hfil$\crcr#2\crcr\sim\crcr}}}

\newcommand{{\SD}}{\rm SD}

\newcommand{{\Mc}}{\mathcal{M}}

\newcommand{\ver}{\mbox{\boldmath${\rm r}$}}

\newcommand{\veP}{\mbox{\boldmath${\rm P}$}}
\newcommand{\vep}{\mbox{\boldmath${\rm p}$}}
\newcommand{\veq}{\mbox{\boldmath${\rm q}$}}
\newcommand{\veQ}{\mbox{\boldmath${\rm Q}$}}
\newcommand{\vez}{\mbox{\boldmath${\rm z}$}}

\newcommand{\veR}{\mbox{\boldmath${\rm R}$}}

\newcommand{\vek}{\mbox{\boldmath${\rm k}$}}

\newcommand{\vexi}{\mbox{\boldmath${\rm \xi}$}}
\newcommand{\veta}{\mbox{\boldmath${\rm \eta}$}}

\newcommand{\vepi}{\mbox{\boldmath${\rm \pi}$}}

\newcommand{\lan}{\langle}
\newcommand{\ran}{\rangle}

\begin{document}
\maketitle
\begin{abstract}
The Lorentz contracted form of the   static wave functions is used to calculate
the  valence parton distributions for mesons and baryons, boosting the rest
frame solutions of the path integral Hamiltonian. It is argued that
nonperturbative parton densities
are due to excited multigluon baryon states. A simple model
is proposed for these states ensuring realistic behavior of
valence and sea quarks and gluon parton densities at $Q^2= 10 (GeV/c)^2$.
 Applying the same model to the proton spin problem
one obtains $\Sigma_3 = 0.18 $ for the same $Q^2$.
 \end{abstract}

 \section{Introduction}

 The partonic model, which allows to express the physical amplitudes in terms
 of quark  and gluon densities, is  now  widely used in many processes \cite{1,2,3}.

 One of basic principles in this approach is the  assumption, that  at high
 momentum $P$ the wave function of a hadron can be represented  as an assembly
 of quasi free partons -- quarks and gluons --  which interact perturbatively and
 are subject to a DGLAP  evolution  \cite{3, 4,5}. E.g. at the first step the
 incident  object (photon) interacts with one of free quarks of the hadron,
 which subsequently emits gluons etc.

 This idea implies, that the  nonperturbative (np) interaction  in   the wave function of
 a very  fast hadron can be neglected as compared  to the   high kinetic energy of
 every parton (quark or gluon). The results of this approach  seem to be  quite
 successful in many  cases and the   whole industry of the parton density
 calculations is now operating  to exploit and  predict experimental data \cite{6,7}.

 From the theoretical point of view this method being intuitively persuasive,
 still lacks rigorous foundations. It is clear, that in  the rest  frame the
 hadron wave function is governed by np interactions, such as
 confinement, and it is not understood how it changes with increasing velocity of
 the hadron.

 One would like to calculate the hadron wave function  in any moving frame and
 demonstrate the resulting transformation  and transition to the pure parton
 picture.

 Recently the possibility of this procedure  was discovered in \cite{8}, where it was
 shown, that the  Lorentz contraction condition on the wave function moving
 with velocity $v$, automatically brings it into  a scaled partonic-like form
 $\psi (p_\bot^{(1)},... p_\bot^{(n)}; x_1,...x_n)$, depending on transverse
 momenta $p_\bot^{(i)}$ and longitudinal momenta  $p_\|^{(i)} = x_i P$.

 The new element of this  valence ``partonic wave function'' is the full scale
 np  interaction, governing dependence of $\psi$ on its
 arguments. In particular, for the simplest $s$-wave meson  wave function one
 obtains the form of valence component $\varphi \left(\vep_\bot,
 M_0\left( x-\frac12\right)\right) =
 \varphi(\sqrt{\vep^2_\bot + M^2_0 (x-\frac12)^2})$ where $M_0$ is the
 meson rest mass and $\varphi (\vek)$ is the Fourier transform of the rest frame
 meson wave function.

  It seems interesting, that the shape of the resulting quark density for this
  purely valence  component (without Regge ladder contribution) partly  resembles the
  known examples (at low $Q^2$ the maximum around $x=\frac12$ for mesons and
  around 1/3 for baryons) however other features, such as admixture of antiquarks,
   gluons and behavior around $x=0$ and 1, are different.
  It is argued in the paper that the main np contribution to all
  parton densities at $x$ not close to 1,is coming from the high excited baryon
  states, and one obtains reasonable results, when the scaled partonic
   formalism is used for the latter.A simple model,called the multihybrid
  baryon model is proposed for the excited states and the results are compared
  to the recent data \cite{6,7} at $Q^2 = 10 $(GeV$/c)^2$
  and can be used as an initial step in the DGLAP evolution.
   Thus the use of the Lorentz contracted form of
  np hadron  might be an interesting step in
  establishing of the new ``nonperturbative parton
  model" and np quark densities. In addition, it may  help to resolve the
  existing difficulties  in the present theory, such as the proton spin problem
  \cite{9}, as we discuss in what follows, see also \cite{9*} for a review.

 Our main result in this paper is the method of constructing the polarized and
 unpolarized parton densities from np wave functions  (n.w.f) of
 any number of quarks and gluons.

 In doing so we exploit the n.w.f. written in the infinite momentum frame,
 which has the  explicit partonic scaled form. In  terms of the relativistic
 Fock-Schwinger Hamiltonian these n.w.f. correspond  to the Fock components
 with definite number of constituents, while the experimental data (in DIS of
 high energy collisions) refer to the  whole Fock columns of these components,
 also developing in time. Therefore to get a full picture one must treat the
 full Fock column wave function, with all relative weights (admixtures) of different
 components.

We show that these admixtures of higher states to the valence component are
Lorentz invariant, however strongly depend on the total energy eigenvalue of
the whole Fock wave function.

E.g.  the proton ground state is well represented by the valence component,
while the proton high  excited states  may contain many   additional gluons,
which are  seen in DIS.

We calculate the valence component of the proton  pdf from the realistic proton
wave function and compare it with DIS  data and model  (QCD sum rules) results.

 At this point  it is important to compare our results with
  the  powerful method of
 light-cone quark models \cite{5a}, see \cite{5b} for a review. In this case
 the Fock components also for polarized parton distributions are derived from
 the light-cone formalism for the wave functions. This approach has given an
 important tool for the analysis of experimental data, see \cite{5c}, and refs.
 therein.

  An essential new ingredient of our approach is the
  implementation of the  np confining properties of the vacuum in
  the boosted instantaneous as well as in the light-cone dynamics.
In the  last case the corresponding Hamiltonian with confinement have  been
derived and studied numerically in \cite{x,xx},   demonstrating the same
confining spectrum for hadrons, as in the instantaneous rest frame. The
light-cone formalism is specifically convenient for the  multihybrids  -- the
string-like objects consisting of $n$ gluons ``sitting'' on  the QCD confining
string, which are possible structures  behind the observed ``ridges'' in
high-energy collisions, as discussed in  \cite{xxx}.

In this way our  approach may  be a  useful  addition to  the well developed
light-cone methods
  in the high-energy and
  high-momentum QCD, introducing string-like objects in the light-cone vacuum,
which are possible in the string theory context.

We start in the paper with the unpolarized parton densities for
the  meson and baryon cases, in the boosted instantaneous
dynamics, \cite{8}, where it was shown that the    np interaction
is  dominant in the valence Fock component, and quarks can be
considered as relativistic and  with suppressed spin degrees of
freedom.(In addition the latter  can be taken  from the
Pauli-Melosh basis approach of the  light-front dynamics
\cite{xxxx}, see \cite{5x} for  a review and references. The
possible use of this formalism for the polarized pdf in our case
is now under consideration.)

To include the excited states we consider the
 sequence of excited multigluon baryons with known form of the c.m. wave functions
and find the corresponding parton form
 for the whole sum, which appears to be close to the recent pdf data \cite{6,7}.

  We also calculate the polarized proton pdf
   for the ground state proton, using the boosted factorized form of the  3q Dirac wave
  functions, and
  show that those are compatible with the
   spin projection criterium. Applying the same
   baryon multihybrid model to the spin projection
  criterium we find that gluons partly
   compensate valence quark spins leaving only about 0.2 of their spin projections.

  The plan  of the paper is as follows. In the next section we write the general  equations
   for the scaled parton distributions from the  nonperturbative (NP) rest
   frame wave functions for two and three partons. In section 3  the QCD Hamiltonian
    and Fock components are discussed both in the rest frame and at nigh $P$.
    We here calculate the proton valence pdf and compare it with known
    examples.

    In section 4 we introduce
     the baryon multihybrid model and calculate pdf's of valence and sea quarks and gluons.

   In section 5 the  proton spin problem is discussed for the relativistic
   proton wave function which ensures the correct $g_A/g_V$ ratio for the ground state
    proton. The contribution of excited states is considered
   in the framework of the baryon multihybrid model
    and the resulting value of $\Sigma_3$ is calculated.

   The last section is devoted to the Summary of the  results and prospectives.

   \section{Parton densities from the Lorentz contracted wave functions}

   The multiparton wave function normalized  in a standard way \cite{2}

\be E(P) \int \prod^N_{i=1} \frac{d^3p_i}{\varepsilon_i} \delta^{(3)} \left(
\veP-\sum^n_{k=1} \vep_k\right)| \psi(p_1,... p_N)|^2 =1 , \varepsilon_i =
\sqrt{\vep^2_i + m^2_i}\label{A.1}\ee   in the limit of large $P$ is written as
  \be \int \prod d^2 p_\bot^{(i)} \frac{dx_i}{x_i}
\delta^{(2)} (\sum^N_{i=1} \vep_\bot^i) \delta (1- \sum^N_{i=1} x_i) |\psi(
p_\bot^{(i)} , x_i)|^2=1 .\label{A.2}\ee

It can be connected to the Lorentz contracted rest frame wave function $\tilde
\varphi_0 (k_\bot^{(1)}, k_\bot^{(2)},... ; k_\|^{(1)} \sqrt{1-v^2},...)$ which
is normalized in the rest frame as \be M_0 \int|\tilde \varphi (\vek^{(1)}
\vek^{(2)}...) |^2 \frac{ d^3\vek^{(1)}}{(2\pi)^3}...
\frac{d^3\vek^{(N)}}{(2\pi)^3} (2\pi)^3  \delta^{(2)} (\sum
\vek_\bot^{(i)})\delta (1-\sum x_i)=1,\label{3}\ee where  $k^{(i)}_\| =
M_0(x_i-\nu_i)$.

The parton distribution in the hadron $h$ is  $D^q_h (x,k_\bot)$, which is
defined as

\be D^q_h (x, k_\bot) = \sum_n \prod_r \frac{d^3k_r}{\varepsilon_r} E_h^2
\delta^{(3)} (P-\sum k_r) | \psi_h^{(n)} (k_r, \lambda_r) |^2\sum_{r(j)}
\delta^{(3)} (k-k_j))\label{4}\ee
$D^q_h (x, k_\bot)$ satisfies the following conditions \cite{2} \be \int d^2
k_\bot dx D_h^{q_j} (x, k_\bot) = N_h^j,\label{5}\ee where $N^j_h$ is the
number of partons  of the type $j$  in the hadron $h$, and normalization
condition

\be \sum_j \int d^2 k_\bot  dx x D_h^{q_j} (x, k_\bot)=1.\label{6}\ee

One can also define the parton density \be D^q_h (x) = \int d^2 k_\bot D^q_h
(x, k_\bot)\label{7}\ee

In terms of the meson  rest frame wave function one can write, taking into
account, that it depends on the relative momentum $\vek$ in the rest frame \be
D^q_M(x) = \frac{M^2_0}{(2 \pi)^3} \left| \tilde \varphi_0^{(2)} \left(k_\bot,
M_0 \left( x-\frac12\right)\right)\right|^2,\label{8}\ee where the meson wave
function is normalized as
$$ \frac{M^2_0}{(2\pi)^3} \int\left| \tilde \varphi_0^{(2)} \left(k_\bot, M_0 \left(
x-\frac12\right)\right)\right|^2 d^2 k_\bot dx=$$

\be=\frac{M_0}{(2\pi)^3} \int\left| \tilde \varphi_0^{(2)} \left(k_\bot,k_\|
\right)\right|^2 d^3 k =1\label{9}\ee

From the condition $\int \left( x-\frac12\right)|\tilde\varphi^{(2)}_0|^2 d^2
k_\bot dx =0$ and normalization condition (\ref{9}) one obtains both relations
(\ref{5}) and (\ref{6}).

For the $3q$ valence wave function of the baryon one can write, e.g. for the
$u$ quark distribution in the  proton (ignoring spin degrees of freedom, see
section 4)

$$u(x, k_\bot) = \int \delta^{(2)} \left(\sum^3_{i=1} k_{\bot i} \right) \prod^3_{i=1} d^2
k_{\bot i} dx_1 dx_2 dx_3 \delta \left(1- \sum x_i\right)\times $$\be \times
\frac{M_0^3}{(2\pi)^3} |\tilde\varphi_0^{(3)}|^2 [(\delta^{(2)} (k_\bot -
k_{\bot 1} ) \delta (x-x_1) + (1\leftrightarrow2))]\label{10}\ee  where $\tilde
\varphi^{(3)}_0$ is $\tilde \varphi^{(3)}_0 \left(k_{\bot 1},... k_{\|
1}^{(0)},... \right).$

The  normalization of the $\tilde \varphi_0^{(3)}$ is \be
\frac{M^3_0}{(2\pi)^6}
  \int \prod^3_{i=1} d^2 p^{(i)} dx_i
   \delta^{(2)} \left(\sum^3_{i=1}
p_{\bot }^{(i)} \right) \delta \left(1- \sum^3_{i=1} x_i\right) \left| \tilde
\varphi_0^{(3)}  (p_{\bot,...,} x_i  \right|^2 =1.\label{11}\ee

In a similar way one defines the $d$ quark distribution, in which case  one
will have one product of $\delta$ functions instead of the sum of two products
in the square brackets in (\ref{10}).

For the 3-particle wave function and Hamiltonian it is convenient to introduce
the total momentum $\veP$ and two relative momenta $\vepi, \veq$ defined as
follows  \cite{10, 11} (we assume for simplicity particles 1 and 2 to be
identical)
$$ \veta = \frac{\vez^{(1)} - \vez^{(2)}}{\sqrt{2}}, ~~ \vexi =
\sqrt{\frac{\omega_3}{2 \omega}} (\vez^{(1)} + \vez^{(2)}- 2 \vez^{(3)})$$ \be
\veR = \frac{1}{\Omega} \sum^3_{i=1} \omega_1 \vez^{(i)},~~ \Omega =
\sum^3_{i=1} \omega_i, ~~ \omega_1=\omega_2=\omega\label{12}\ee

$$ \veP = \frac{\partial}{i\partial \veR}, ~~ \veq = \frac{\partial}{i\partial
\vexi}, ~~  \vepi = \frac{\partial}{i\partial\veta}$$

\be H = \frac{\veP^2}{2\Omega} + \frac{\veq^2+\vepi^2}{2\omega} + \sum^3_{i=1}
\frac{m^2_i+\omega^2_i}{2\omega_i} + V (\veta, \vexi).\label{13}\ee

In terms of individual momenta $\vep^{(i)}= \frac{1\partial}{i\partial
\vez^{(i)}}$ one has the following connection
$$ \vep^{(1)} =\frac{\omega}{\Omega} \veP +  \sqrt{ \frac{ \omega_3}{2\Omega}}
\veq -\frac{\vepi}{\sqrt{2}}$$

$$ \vep^{(2)} =\frac{\omega}{\Omega} \veP +  \sqrt{ \frac{ \omega_3}{2\Omega}}
\veq + \frac{\vepi}{\sqrt{2}}$$ \be \vep^{(3)} =\frac{\omega_3}{\Omega} \veP -
\sqrt{ \frac{ 2\omega_3}{ \Omega}} \veq \label{14}\ee Note, that $\omega_i$ are
found from the stationary point analysis of $M_n (\omega_1, \omega_2,
\omega_3)$ the eigenvalue of $H$, namely from the relations \be
\left.\frac{\partial M_n ( \{\omega_i \})}{\partial \omega_k} \right|_{\omega_k
= \omega_k^{(0)}}=0.\label{15}\ee

Since $\tilde \varphi_0^{(3)}$ is the Fourier transform of the rest frame wave
function, we shall use finally the values of $\omega_i^{(0)}$ obtained in the
rest frame.

Using (\ref{14}) one can write in (\ref{11}) \be \delta^{(2)} (\sum^3_{i=1}
\vep_\bot^{(i)}) \prod d^2 \vep_\bot^{(i)}  = \frac{\omega_3}{\Omega} d^2
\veP_\bot d^2 q_\bot d^2 \vepi_\bot d^2 \vepi_\bot \delta(\veP_\bot)=
\frac{\omega_3}{\Omega} d^2\veq_\bot d^2\vepi_\bot.\label{16}\ee

To find the arguments $p_{\| i}^{(0)} (x_i)$ in (\ref{10}), (\ref{11}), which
are the longitudinal components of parton momenta in the rest frame, we use the
Lorentz connection of these to the momenta in the moving frame \be p_{\| i} =
\frac{ p_{\| i}^{(0)} + v\bar \omega^{(0)}_i}{\sqrt{1-v^2}} = P
x_i,\label{17}\ee which yields \be p_{\| i}^{(0)} = \left(x_i - \frac{\bar
\omega_i^{(0)}}{M_0}\right) M_0.\label{18}\ee In what follows we shall accept
for simplicity the relation for massless quark $ \bar \omega_i^{(0)}\equiv
\bar\omega^{(0)}, ~~ i=1,2,3$ and   the relation for the total mass in the case
of free quarks, which holds  approximately true in the interacting case $ M_0 =
\sum^3_{i=1} \bar \omega_i^{(0)} = 3 \bar \omega^{(0)},$ hence $p_{i \| }^{(0)}
= \left( x_i -\frac13 \right) M_0$. For the relative momenta one obtains \be
q_\|^{(0)} = \frac{M_0}{\sqrt{6}} (x_1+ x_2 - 2 x_3), ~~ \pi_{\|}^{(0)} =
\frac{M_0 (x_2-x_1)}{\sqrt{2}}.\label{19}\ee One can rewrite the normalization
condition (\ref{11}) as \be 1= \frac{M^3_0}{(2\pi)^6} \cdot \frac13 \int d^2
\veq_\bot d^2 \vepi_\bot dx_1 dx_2 dx_3 \delta \left( 1- \sum^3_{i=1}
x_i\right) \left| \tilde \varphi_0^{(3)} (\veq_\bot, \vepi_\bot, q_\|^{(0)},
\pi_\|^{(0)} \right|^2\label{20}\ee and Eq. (\ref{10}) acquires the form $$
u(x, k_\bot) = \frac{4 M^3_0}{(2\pi)^3} \int d^2 \vepi_\bot \int^1_0 $$ \be d
x_2 \left| \tilde \varphi^{(3)}_0 \left(\sqrt{3} \vepi_\bot + \sqrt{6}
\vek_\bot; \vepi_\bot; \sqrt{\frac32} M_0 \left(x+x_2-\frac23\right);\right.
\left.\left(\frac{M_0 (x_2-x)}{\sqrt{2}}\right)\right)\right|^2.\label{21}\ee

It is known, however, that the nucleon wave function can be
expanded in the series of hyperspherical harmonics \cite{11}, and
the leading term accounts for $\sim 90\%$ of the normalization
condition. This means, that $\tilde \varphi_0^{(3)}$ can be
considered as the function of \be \veQ^2 = \vepi^2 + \veq^2 =
\vepi^2_\bot + \veq^2_\bot + (\pi_\|^{(0)})^2 + (q_\|^{(0)})^2=
\vepi^2_\bot + \veq^2_\bot + \frac{M_0^2}{3} \sum_{i>j}
(x_i-x_j)^2.\label{22}\ee

As a result the argument  of $\tilde \varphi_0^{(3)}$ in (\ref{21}) can be
written as \be \tilde \varphi^{(3)}_0 (\veQ^2) \to \tilde \varphi^{(3)}_0 ( 4
\vepi^2_\bot + 6 \vek^2_\bot + 6 \sqrt{2} \vepi_\bot \vek_\bot +
\frac{2M_0^2}{3} f(x, x_2))\label{23}\ee where \be f(x, x_2) = 1+ 3x^2+ 3x^2_2
+ 3xx_2 -3x-3x_2= 3 \left( \left( x_2 - \frac{1-x}{2}\right)^2 +
c(x)\right),\label{24}\ee$$ ~~ c(x) =\frac{9x^2-6x+1}{12}.$$

It is conceivable, that after the integration over $d^2\vepi_\bot$ and $dx_2$
the result will depend mostly on $c(x)$, which has the minimum at $x=\frac13$
and hence $u(x, k_\bot)$ as a function of $x$ would have a maximum  around that
point.

This result is close to that obtained for the valence quark density in
\cite{12} and has a  form similar  to the DIS data \cite{7},  \cite{13}  at low
$Q^2$.   In  Fig. 1  we compare our results with those of \cite{12}. We shall
argue however, that the DIS data refer to the high excited baryon state, and
hence could differ  from the ground state
 proton case. In section 4 we provide a simple model of excited states
which yields reasonable results at $Q^2 = 10$ (GeV/c)$^2$
 and can be used for the DGLAP evolution.

\begin{figure}[t]
  \hspace{2cm}
\includegraphics[width=2.6in]{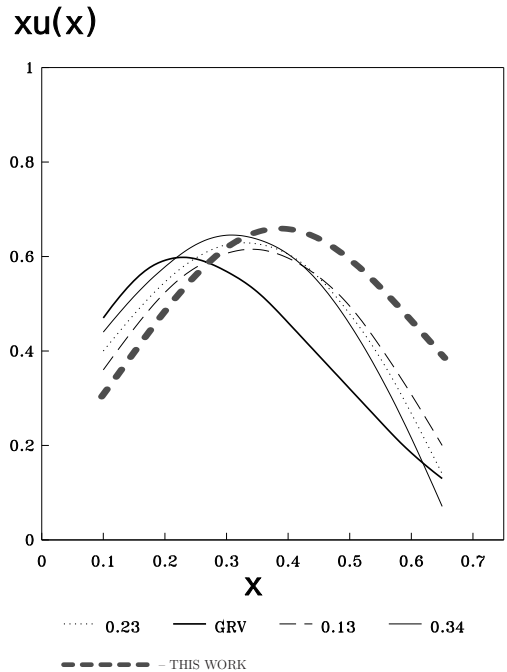}
 \caption{The valence parton distribution
 in proton (thick dashed line) in comparison
 with the  results of \cite{12} for fixed values of quark condensate,
  and results of the second reference \cite{7}, -- thick solid line (GRV).}
\end{figure}

 Note, that our hyperspherical approximation for the proton wave function used
 in Fig.1, thick dashed line, is applicable around $x=1/3$ and not  suitable
 for the region of small 1-x, where quark carries all proton momentum and one
 expects the behavior $u(x) \sim (1-x)^2$ in the free parton case.

\section{The QCD Hamiltonian and Fock components}

We assume in this section that one can construct a QCD Hamiltonian, which
provides eigenvalues and eigenfunction for all Fock components, e.g. in case of
a baryon state, the pure valence state $(qqq)$, also with any number of
additional gluons $(qqqg), (qqqgg),...,$ which are   actually hybrid states,
and with additional $q\bar q$ pairs:$(qqq(q\bar q))$ etc.

One example of such Hamiltonian is provided by the path integral Hamiltonian
  derived in the  framework of the Fock-Feynman-Schwinger Representation
(FFSR) \cite{14}, and developed further in \cite{15,16}. It was used for mesons
\cite{17}, baryons \cite{11,18}, hybrids \cite{19} and glueballs \cite{20}.
yielding in all cases spectra in good agreement with experimental and lattice
data.

The full Fock matrix Hamiltonian in this case consists of the diagonal elements
$H_{nn}^{(0)}$ for each $n$-th Fock component and of the nondiaginal elements
$H_{nk}$, $n\neq k$, which are actually elements of the interaction vertices
$V_{nk}$. In what follows we are using the line of reasoning from \cite{21,22}
and start with the center of mass frame, $\veP=0$. Note, that we have the
instantaneous dynamics, so that all operators $\hat H = \hat H^{(0)}+\hat V$ do
not depend on time, and $\hat V$ acts at the instantaneous moment of creation
or destruction of an additional particle.

 In Fig.2 it is shown, how subsequent Fock components appear in the meson
 Green's function, where additional gluons and a $ q\bar q$ pair are created by
  the interaction $V_{nk}$.

\begin{figure}[!htb]
\begin{center}
\includegraphics[angle=0,width=9 cm]{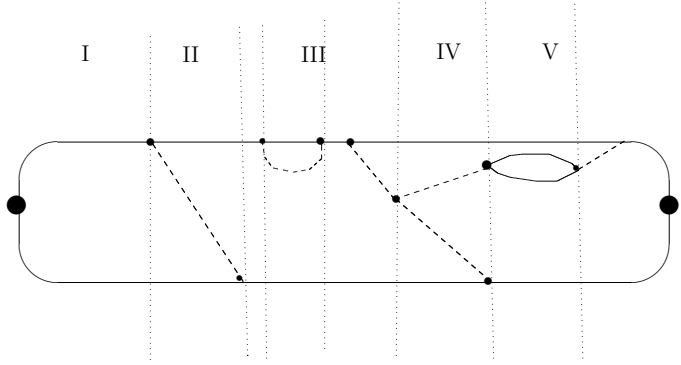}
\caption{ The  meson Green's function, containing the Fock components: $q\bar
q$ (sector I), $q\bar qg$ (sector II and sector III), $q\bar q gg$ (sector IV)
and  $q\bar q q \bar q $ (sector V). Note, that all surface between external
solid lines is filled in by the string world sheet, except for the narrow  gap
between $q\bar q$ lines in sector V }
\end{center}
\end{figure}

One can simplify, as in \cite{21}, the structure of $\hat H$, and taking the
limit of large $N_c$, so  that  an additional $q\bar q$ pair ``costs'' a factor
of $1/N_c$, while an additional glueball gives $1/N_c^2$, so that all Fock
components reduce to  the pure valence  states and its hybrid excitations.

 The basic  matrix equation is  simply

\be
 \hat H\Psi_N=(\hat H^{(0)}+\hat V) \Psi_N=E_N\Psi_N,\label{25}\ee
 and the Fock column
$\Psi_N\{P,\xi,n\}$ has quantum numbers $N=0,1,2$... of the ground  and excited
states,  $n$ refers to the Fock column number, and $\xi$ denotes  additional
internal numbers for a given type of excitation. In the diagonal approximation
$
 \Psi_N\to \Psi_{N\{P, \xi,n\}}$ with $n=n^{(0)}$. Note, that for $N_c\to \infty$
 and baryon number $B=1$ one has only discrete spectrum of a valence and hybrid states.
  In this case the eigenvalues for $\veP\neq 0$ are simply
 \be
 E_N^{(0)} = E_n^{(0)} (P)
 =\sqrt{\veP^2+M^2_{n\{k\}}}.\label{26}\ee
  and the eigenfunctions are $\psi_n (P, \xi, k) \equiv \psi_{n,k}$, where
  $\xi, k$ comprise radial and angular $(k)$, as well as additional $(\xi)$
  quantum numbers. The set $\psi_{n\{k\}}$ can be used to expand the total wave
  function $\Psi_N$, (discrete spectrum at $N_c\to \infty$)
  \be
  \Psi_N=\sum_{m\{k\}}c^N_{m\{k\}}\psi_{m\{k\}},~~
\int\Psi^+_N\Psi_M d\Gamma = \sum_{m\{k\}} c^N_{m\{k\}} c^M_{m\{k\}} =
\delta_{NM} .\label{27}\ee
   As in \cite{21,22}, we use  the orthonormality  condition
  \be
  \int\psi^+_{m\{k\}}\psi_{n\{p\}} d\Gamma
  =\delta_{mn}\delta_{\{k\}\{p\}}\label{28}\ee
   to find the equation for $c$ and $\Psi$

  \be c^N_{n\{p\}} (E_N-E^{(0)}_{n\{p\}}) = \sum_{m\{k\}} c^N_{m\{
  k\}}V_{n\{p\}, m\{k\}}\label{29}
  \ee
  with
  \be
  V_{n\{p\}, m\{k\}}= \int \psi^+_{n\{p\}}\hat V \psi_{m\{k\}}
  d\Gamma.\label{30}\ee to the first order in $V$ one has
 \be
 C^{N(1)}_{n\{p\}}=
 \frac{V_{n\{p\},\nu\{\kappa\}}}{E^{(0)}_{\nu\{\kappa\}}-E^{(0)}_{n\{p\}}} \label{31}\ee
and for the high Fock component with $l$ gluons in addition to the valence
quarks one has in the lowest approximation

$$
 C^{N(
\nu\{\kappa\})}_{\nu+l,\{k\}}= \sum_{\{k_1\}...\{k_l\}}
\frac{V_{\nu+l\{k\},\nu+l-1\{k_1\}}}{E^{(0)}_{\nu\{\kappa\}} -
E^{(0)}_{\nu+l\{k\}}}\frac{V_{\nu+l-1\{k_1\},\nu+l-2\{k_2\}}}{E^{(0)}_{\nu\{\kappa\}}
- E^{(0)}_{\nu+l-1\{k_1\}}}...$$ \be \frac
{V_{\nu+1\{k_l\},\nu\{\kappa\}}}{E^{(0)}_{\nu\{\kappa\}} -
E^{(0)}_{\nu+1\{k_l\}}}+O(V^{l+2}).\label{32}\ee

Using (\ref{27}), (\ref{28}), one can  obtain the equality \be
\sum_{n\{k\}}\left|C^N_{n\{k\}}\right|^2=1,\label{33}\ee and each state $N$ can
be characterized by the sequence
$$\{|C^N_0|^2,  |C_{1\{k\}}^N|^2,...\}\equiv \{C^N\}.$$

Note, that $\hat V$ is $O(g)$, and hence in the limit $g\to 0$ one has the
unmixed states $\{1,0,0,...\}, \{0,1,0,...\}$ etc., while the inclusion of
$\hat V$ starts the ``evolution'' of the basic state along the chain of
neighboring states. This  can be done in principle   in accordance  with the
DGLAP evolution equation, to be written  in terms of parton distribution
functions (pdf), i.e. in terms of $|\psi_{n\{k\}}|^2$ integrated over all pairs
$p_\bot^{(i)}$ and $x_i$, except one, $\vep_\bot^{(i)} \equiv \vek_\bot,
x_1\equiv x$.

Thus the nonperturbative $N_c\to \infty$ limit of the
evolution is given by (\ref{29}), (\ref{31}). At this point it is important to
stress, that sequences $\{C^N\}$ can be completely different for the baryon
ground state with $E_0=m_p$, and the highly excited baryon state with a large
c.m. energy $E_N$. In our case $(N_c=\infty)$ this refers to the discrete
spectrum, while allowing for the $q\bar q$ pairs one has a continuous spectrum.
Indeed, as was estimated in \cite{22,23} for the meson-hybrid mixing
coefficient $V_{Mh}\sim g\cdot 0.08$ GeV  and \be
 C_{Mh} =\frac{V_{Mh}}{E^{(0)}_M-E^{(0)}_h} =\frac{V_{Mh}}{\Delta
 M_{Mh}},~~  \Delta M_{Mh} \sim O (1 ~{\rm  GeV})\label{34}\ee
 which gives the hybrid admixture  $ |C_{Mh}|^2\approx O(1\%)$.
As shown in \cite{21,22} the addition of one gluon to the  hybrid state
``costs'' around 1 GeV, hence  multigluon states contribute very little to the
ground state wave function.
 For the  high excited states the
 denominator in (\ref{34}) can be much smaller and the total number of mixing
 states grows, so that the gluon admixture should grow substantially.

 This is exactly what happens in DIS. Indeed, the c.m. energy of the baryon
 state, with initial momentum $p$, exited by the incident virtual $\gamma$ or
 $W,Z$ with momentum $q$, is very high in the Bjorken limit.
 $$ s=m^2_B + 2\nu(1-x), ~~ x = Q^2/2\nu, ~~ s-m^2_B = 2\nu(1-x)\gg m_B^2.$$

 In Fig.~3  we  show schematically how the excited baryon state emerges in DIS.

\begin{figure}[!htb]
\begin{center}
\includegraphics[angle=0,width=9 cm]{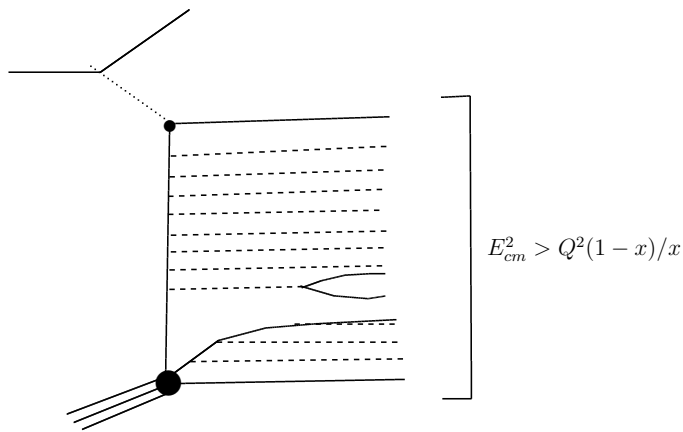}
\caption{The excited baryon state created in DIS has the excitation energy in
the rest frame  $E_{CM} = \sqrt{M^2_B + \frac{Q^2(1-x)}{x}}$,  which is  much
larger than $M_B$ and the boost momentum $Q$ for small $x$}.
\end{center}
\end{figure}

 It is actually the baryon state with the c.m. energy $E_{cm} = \sqrt{s}$,
 which is tested in DIS and the resulting  pdf  refer to this $E_{cm}$, which
 is close to $m^2_B$ only at the end point $x=1$. Therefore one should expect,
 that at growing $s=m^2_B+ \frac{Q^2}{x} (1-x)$ the admixture of gluons should
 grow fast, since the coefficients (\ref{32}) increase with energy, or
 equivalently, DGLAP evolution at high $Q^2$ produce a large gluon component,
 see e.g. \cite{6,13}.  We stress again., that this fact refers not to the ground state
 described above. For the ground state baryon (or meson) the sequence $\{C^N\}$
 is fast decreasing and is given by the  rest frame wave functions,
 described in the previous section.

 One of the consequences of this discussion is the  possible resolution of the proton
 spin problem  in the  next section (see \cite{9} for discussion and references).
  Indeed, insertion in the
 proton wave function the polarized parton distributions, obtained from the
 highly excited baryon states, results in the high admixture of the antiquark
 and gluon components  (which are suppressed in the genuine proton wave
 function). As a result the contribution of the valence quarks is very small,
 and one faces the strange picture of the almost quarkless proton. If instead
 one uses the pdf of the multigluon baryon, described in the next section,
 this discrepancy disappears, as we show  in the section 5.

 One should stress in addition, that the virtual photon in DIS is able to
 transfer any amount of the angular momentum (similarly to the process  of the
 electroexcitation of nuclei), so that the excited baryon can have any half
 integer  spin, suppressing in this way the contribution of the $s=1/2$.

We now turn to the boosted form of the hadron wave function. We assume, as in
\cite{8}, that the boost acts on the spacial wave function $\Psi_N$ as the
Lorentz contraction, while the interaction term $\hat V$  behaves as $L\hat V =
CV, ~~ C_0 = \sqrt{1-v^2}.$

This becomes  clear from (\ref{25}), if one writes the Hamiltonian in
(\ref{25}) in the  off-shell form as in (\ref{13}) \be \hat H^{(0)} =
\frac{\veP^2+\Omega^2}{2\Omega} + \sum^N_{i=1} \frac{m^2_i + \omega^2_i
+\vep^2_i }{2\omega_i} + \frac{C(V_0 + \Delta V)\bar M}{2\Omega},\label{35}\ee
where we have splitted the interaction $\hat V = V_0 + \Delta V$ into a
diagonal and nondiagonal parts in the number $n$ of constituents.

As a result  $$ E_n^{(0)} = \sqrt{ \veP^2 + (M_n^{(0)})^2} \simeq P+
\frac{(M_n^{(0)})^2}{2P}+...$$ and \be C_{n\{P\}|^N\{1\}}=
\frac{CV^{(0)}_{n,n \pm 1}}{E^{(0)}_{n\pm 1}- E_n^{(0)}}\simeq \frac{C_0P V_{n,
n\pm 1}^{(0)}}{(M^{(0)}_{n\pm 1})^2 - (M_n^{(0)})^2} \simeq \frac{M_n^{(0)}
V_{n, n+1}^{(0)}}{(M^{(0)}_{n\pm 1})^2 - (M_n^{(0)})^2}.\label{36}\ee Hence the
set $\{C^N\}$ (\ref{34}) is boost invariant and the nucleon mixing contents
does not change, when going from the rest frame to the infinite momentum frame.
The same can be said about all excited baryon states, which are measured using
4-point amplitudes, as it is done in DIS, and the high excited energy states
(about tens of GeV on average) are very different from  the ground state
nucleon both in the c.m. and infinite momentum frames.  In the DIS partonic set
$\{C^N\}$ one has much higher admixture of antiquark and gluon (hybrid)
components, as compared to the ground state nucleon partonic set. This possibly
explains the long standing proton spin puzzle \cite{9}, where the use of the
polarized DIS data yields high antiquark admixture, cancelling the valence
quark contribution, and high gluon contribution.

At this point it is  essential to note, that the very idea, that the high
excited hadron can be represented by an assembly of almost free partons seems
to be reasonable, but the idea, that the fast moving ground state hadron can be
represented by  the same  set of free partons from our point of view has no
foundations.

It is important, that  in DIS we talk about the high-excited baryon Green's
 function and  parton distributions, which is especially evident, when one considers the
Regge exchange contribution to the parton densities.

In   some cases, as in DIS, quantum numbers of this high-excited object can be the
same, as for a nucleon, but we stress, that the ground state nucleon has
 little  to do with this object quark densities, while  it is likely, that the
high excited hadron can simulate the observed quark distribution. It is even
likely, that at very high excitation   the nonperturbative contributions  can
be treated as the initial state to the evolution of partons subject to  the
DGLAP or BFKL  procedure, see e.g. \cite{24}.

\section{Nonperturbative model of DIS parton distributions in proton }

The purpose of this section is to provide a simple model of the np DIS
process,which can demonstrate a large gluon (and sea quark) component at small
$x$, found in the data \cite{6,7}, as compared to the valence component. To
this end we consider the region of $x$ not close to 1 and $Q^2$ around 10
(GeV/c)$ ^2$ and we expect that the main part of all products in DIS is the
result of the set of high excited baryonic states, which in the limit $N_c\to
\infty$ one can associate with the multihybrid states \cite{xxx}.

For the multihybrid state, which is the basic component in the
large $N_c$ limit, the gluons, depicted as broken lines in Fig. 3,
are connected by the QCD string,and the corresponding structure is
actually the excited gluon string, where excitations have the
particle-like form,as it is known in hybrids \cite{19,21,22}.

We shall use the parton distribution function (pdf) $D^i_N (x)$
 as in (\ref{4}), and we shall try the factorization ansatz
for the wave function of the N- multihybrid, consisting of 3
quarks and N gluons

\be |\Psi^N_B ( k_{\bot i}, k_{\| i})|^2 =\prod^{N +3}_{i=1} |\varphi_i
(k_{\bot i},k_{\| i})|^2, \label{37} \ee where $\varphi_i$ can be written as
$\varphi_i ((k_\bot i)^2, (k_\| i)^2)$ and according to \cite{8} the parton
form is obtained as follows \be \varphi_i (k_\bot,k_\|) = \varphi_i ((k_\bot)^2
+ M_N^2 (x_i - \nu_i)^2) .\label{38} \ee

The Hamiltonian for the 3q multihybrid state, consisting of 3
quarks and N gluons, "sitting " on the strings connecting quarks
to the string junction, can be written in full analogy to the case
of the $q\bar q$ multihybrid state studied in \cite{19,21,22}, where
quarks and gluons entered additively. Therefore one can write the
resulting total energy as

\be M_N = 3\omega + N m, 3\omega = M_p, \label{39} \ee where $\omega$ and $m$
are c.m. energies of the quark and gluon respectively. In the wave function
(\ref{39})the term $\nu_i$ according to \cite{8} can be written as

\be \nu_q = \omega/M_N ,~~ \nu_g = m/M_N. \label{40} \ee

In the large N limit one can write $D_N^i(x)$ as follows

\be D_N^i(x)= f_i^{(N)}(M_N |x-\nu_i|),~ i= q,g \label{41} \ee
where $f_i(y)$ is subject to the normalization condition $$
\int^1_0 f_i(x) dx = 1. $$ In the particular case of the Gaussian
wave function, which was shown in \cite{xxx,22} to be a good
approximation for the multihybrid wave function, one has

\be f_i^{(N)}(M_N |x-\nu|) =
 \xi_i^{(N)} \frac{M_N}{\kappa} \exp( -\frac{M_N^2}{\kappa^2} (x-\nu_i)^2),
\label{42} \ee
and $\xi_i$ is defined by the normalization condition of $f_i$.

We are now in a position to construct pdf
 for quarks and gluons, which can serve as a nonperturbative
input at some $Q^2$ and being
 evolved by the DGLAP mechanism. One of not still understood features of
the standard theory is the pdf
 behavior at very small
  $x < 10^{-3}$, where $x g(x)$ is diverging (seemingly as $x^{-0.5}$,
   while $x u(x)$ is behaving approximately as $x^{0.5}$ \cite{6,7}.

In what follows we shall be mostly interested in this region of $x$, leaving
the region near $x = 1$ for the lack of space, and we shall show, that our
model can explain this dependence at $Q^2 = 10 $(GeV/c)$^2$, while gluon
evolution and quark pair production can explain the behavior at larger $Q^2$.
To this end we assume that the main step in the DIS process is the creation of
the multihybrid baryon sequence of states with coefficients $C_N$, satisfying
the orthonormality condition (\ref{33})

\be | C_0|^2 + \sum^\infty_{N=1} |C_N|^2 = 1. \label{43} \ee

Note,that $|C_0|^2$ gives the prbability of the
 pure baryon state (without gluons, but with radial and orbital excitations),
 which we combine into one baryon state, and similarly for $|C_N|^2$ which include
possible excitations. We shall have in mind the limit of large
$N_c$, which allows for gluon multiplication, but forbids in the
lowest order the $q \bar q$ creation by gluons.

As a result one can write

\be g(x) = \sum^\infty_{N=1} N f_g^{(N)}(x) ,~~ u_v(x) = 2
\sum^\infty_{N+1} f_q^{(N)}(x) + 2 f_q^{(0)}(x). \label{44} \ee

Inserting $f_i^{(N)}$ from (\ref{43}) one obtains

\be u_v(x) = u_0(x) + u_g(x), \label{45} \ee where $u_0(x)$ and
$u_g(x)$ are

\be u_0(x) = 2 |C_0|^2 M_p \frac{\xi_q^{(0)}}{\kappa}
 \exp{(-\frac{M_p^2}{\kappa^2} (x-1/3)^2)}  \label{46} \ee

\be u_g(x) = 2 \sum^\infty_{N=1} |C_N|^2 \xi_q^{(N)} (M_p + m N)
 \exp{( -\frac{M_N^2}{\kappa^2} (x-\nu_q)^2)},
\label{47} \ee

 \be g(x) = \sum^\infty_{N=1} |C_N|^2 |\xi_g^{(N)} N \frac{M_p + N m}{\kappa}
 \exp\left\{\left(- \frac{M_N^2}{\kappa^2}
   (x-\nu_g)^2\right)\right\}.  \label{48} \ee

Here $\xi_{q,g}$ are defined by normalization

\be \xi_q^{-1} = \int^\frac{M_N}{\kappa}_0 d y \exp{(-(y-\omega/\kappa)^2)} \label{49} \ee

\be \xi_g^{-1} = \int^{\frac{M_N}{\kappa}}_0 d y
\exp{(-(y-m/\kappa)^2)}. \label{50} \ee

 At this point, having in mind that the effective
  region of $N$ in the sums over $N$ is $N\gg 1$,
   it is useful to replace the sums by the
   integrals over $N$ and change variables as follows

 \be y = \frac{N m x}{\kappa},~~ N = \frac{y \kappa}{m x} .\label{51}\ee

  To obtain the resulting
   difference in the  behavior of $u_v(x), g(x)$ we assume the ansatz

  \be |C_N|^2 = N^{-3/2} |\bar c|^2,~~ N \gg 1 .\label{52}\ee

As a result we are left with the parameters
 $m, \kappa, |c_0|^2, |\bar c|^2$, the latter satisfying
the condition

\be |c_0|^2 + |\bar c|^2 \zeta(3/2) = 1, \zeta(3/2) = 2.62
,\label{53} \ee and $u_v(x), g(x)$ assume the form

 \be u_0(x) = 2 M_p |c_0|^2 \xi_q^{(0)}
 \kappa^{-1} \exp{(-\frac{M_p^2}{\kappa^2} (x-1/3)^2)}, \label{54} \ee

 \be u_g(x) = 2 |\bar c|^2 \xi_q (\sqrt{\frac{m}{\kappa x}} I_q(-1/2) +
 \frac{M_p}{\kappa} \sqrt{\frac{m x}{\kappa}}  I_q(-3/2))
,\label{55} \ee

\be  g(x) =|\bar c|^2 \xi_g (\sqrt{\frac{M_p^2}{\kappa m x}} I_g(-1/2) +
   \sqrt{\frac{\kappa}{m}} x^{-3/2} I_g(1/2)) ,\label{56} \ee
where notation is used

   \be I_q(n) = \int^\infty_{ m x/\kappa}
   d y y^n \exp{(-(\frac{M_p x -\omega}{\kappa} + y)^2)} ,\label{57} \ee

 \be  I_g(n) = \int^\infty_{m x/\kappa} d y y^n \exp{(-(\frac{M_p x -m}{\kappa} + y)^2)}
  , \label{58} \ee
  and we have taken the asymptotic values

   \be \xi_q = (1.63)^{-1}, \xi_g = (1.77)^{-1}, \xi_q^{(0)} = (1.48)^{-1}  .\label{59} \ee

To fix the retaining parameters we use the information on the $q\bar q g$
hybrid state \cite{19,22}, giving the gluon energy $m = 0.7$ GeV  and $\kappa =
0.36$ GeV, since $M_p = 0.94$ GeV  we take $\omega = \kappa = m/2 = 0.313$ GeV,
leaving the problem without free parameters. At small $x \ll 0.1$ one has
$I_g(n) \simeq 2^n \sqrt\pi$ and $g(x)$ acquires the form

\be g(x) = |\bar c|^2 (1.5 x^{-1/2} + x^{-3/2}) ,\label{60} \ee
while at $x = 2/3$ one obtains $$xg(x) = 0.11 |\bar c|^2$$.
  For $u(x)$ one has in the small $x$ region,
 $x \ll 0.1$,

 \be u(x) = 1.5 |c_0|^2 + |\bar c|^2 (2.82 x^{-1/2} + 2.70) ,\label{61} \ee
while at $x = 1/3$ one has

 \be u(x) = 4.05 |c_0|^2 + 1.74 |\bar c|^2 \label{62} \ee.

  We can now compare our results with the data from (\cite{6,7}),
   which we call for brevity the PDG data,
  see especially the Fig.4 of the first reference in(\cite{7}).
   It is evident from , that
  the best solution can be obtained for $c_0 = 0$,
  and hence from (\ref{53}) one obtains $|\bar c|^2 =(2.62)^{-1}$.
   As a result we quote predictions of our multihybrid model
    in comparison with the PDG data. For $x u_v(x)$ and a
     sequence of $x$ values $x = 10^{-3};10^{-2};10^{-1};0.5$ we obtain from (\ref{61},\ref{62})
  $0.037;0.123;0.46;0.38$ versus PDG data $0.03-0.07;0.16-0.19;0.52;0.23$.
  At the same time for the gluon pdf $x g(x)$
   at the sequence of points $x = 10^{-2};10^{-1};2/3$ our model predicts
  $3.87;1.38;0.042$ to be compared with the PDG data for the same $x$
   values $5;1.5;0.05$. One can see, that
  the qualitative agreement is reasonable, also taking
   into account that our initial values are subject to
  the subsequent evolution.

Till now we studied the pdf of valence quarks and gluons
 to stress the basic difference between their
   behavior near $x=0$, which cannot be explained by the
    perturbative QCD evolution, but should be inserted
   beforehand. Note, that
    in our model this difference occurs naturally due to additional factor N in (\ref{48}),
     i.e. due to many gluons with the fixed number of quarks.

To calculate sea quark pdf one can  use the DGLAP evolution
equations,which to the lowest order yield

\be \frac{Q^2 \partial \bar u(x,Q^2)}{\partial Q^2} =
\frac{\alpha_s(Q^2)}{2 \pi} \int^1_x d y \frac{(x/y)^2 +
(1-x/y)^2}{2 y} g(y,Q^2). \label{63} \ee

Considering the interval $1 GeV^2 < Q^2 < 10 GeV^2$
one can integrate (\ref{63}) with the resulting
estimate

\be \bar u(x,Q^2 = 10 GeV^2) \simeq 0.04 g(x,Q^2 = 10 GeV^2),
\label{64} \ee which roughly agrees with the PDG data, yielding
for the same ratio the value $0.053$. Thus one can see, that the
model can explain the basic difference between valence quarks and
gluon pdf and also provides a reasonable estimate for the sea
quark parton densities. To go to higher $Q^2$ one
 should use the standard DGLAP formalism providing the growth of
 $g(x)$ and $\bar u(x)$ with $Q^2$
 due to gluon proliferation while the valence quark pdf
 changes only a little.We have not studied above
 the behavior near $x = 1$, which needs consideration
 of the almost elastic collision with the power
 counting methods.

\section{Polarized parton distributions and $g_A/g_V$ in the proton }

   In this section we shall derive the polarized parton distributions and
   nucleon  axial charges, starting with the  rest frame nucleon wave function.
    In section 2 we have defined the unpolarized parton distributions  using
    the nucleon wave function, which has a simple structure without lower Dirac
    components. However for a reliable description of spin degrees of freedom
    and axial charge this is not enough and one must use  the full 4 component
    Dirac structure of every  quark. To this end we use the the decomposition
    of the 3q wave function in the products of Dirac quark bispinors -- the
    so-called Dirac orbital model \cite{25,27} and keep for simplicity only the
    first  dominant term in the sum over spin, isospin and angular momenta.
    \be \Psi(\ver_1, \ver_2, \ver_3) = \sum_{\{ n_i\}} \prod^3_{i=1}\psi_{ n_i}
    (\ver_i) C_{n_1n_2n_3}.\label{65}\ee
    In the momentum space one can write for the nucleon
    \be \tilde \Psi_N (\vep_1, \vep_2, \vep_3) = \sum_{\alpha_i} \prod^3_{i=1}
    \phi_{\alpha_i} (\vep_i) C_{\alpha_1\alpha_2 \alpha_3}, \label{66}\ee
    where  $\sum\vep_i=0$ and $\alpha_i$ stand for  spin-isospin and the
    angular momentum variables. As it is known from the actual calculations
    \cite{11,18, 28} the dominant contribution in the 3 body nucleon wave
    function is given  by the symmetric in quarks component, which can be
    written for the proton with the spin up as
    \be
    \Psi_p \equiv (p\uparrow\ran = \frac{1}{\sqrt{18}} [-2 ( | u\uparrow
    u\uparrow d\downarrow\ran  +{\rm  perm} ) + (|u\uparrow u\downarrow d\uparrow\ran
    +{\rm perm})]\label{67}\ee
    and perm implies the permutation of the quark positions  in the  the
    triade, while each of quark functions is a Dirac bispinor,  viz.
    \be \chi_\uparrow (r, \theta, \phi) = \frac{1}{r} \left(\begin{array}{l}
    G(r)\Omega_{\frac12 0 \frac12}\\F(r) \Omega_{\frac12
    1\frac12}\end{array}\right),\label{68}\ee
    normalized as \be \int^\infty_0 (G^2(r) + F^2(r)) dr =1.\label{69}\ee
    Now one can define the proton axial charge \cite{27}

    \be g_A = \lan p\uparrow | \hat u^+ \Sigma_3 \hat u - \hat d^+ \Sigma_3 \hat d
    |p\uparrow\ran, \label{70}\ee
    where   $\Sigma_3= \left(\begin{array}{ll}
    \sigma_3&0\\0&\sigma_3\end{array}\right)$, and one obtains
    \be  g_A=\frac43 \left\lan \chi_\uparrow|\Sigma_3| \chi_\uparrow\right\ran -\frac13\left\lan \chi_\downarrow|\Sigma_3| \chi_\downarrow
     \right\ran=
     \frac53\int^\infty_0 \left(G^2(r)-\frac13 F^2(r)\right) dr= \frac53 \left( 1-\frac43 \eta\right),\label{71}\ee
with $\eta = \int^\infty_0 F^2 (r) dr$. As was shown in \cite{27}, the
calculation of $\eta$ for $\alpha_s=0.39$ yields $g_A=1.27$ in good  agreement
with experiment \cite{29,30}.

In a similar way one can define the proton spin projection, which has a general
form \cite{9}, which is made gauge and boost invariant, using Fock space
Hamiltonian solutions,  \be J_3 =\frac12 \Sigma_3 + \Delta L_q + \Delta G +
\Delta L_g = \frac12,\label{72}\ee where for all operators one should take
matrix element between the states $|\vep \uparrow\ran$, and $\Delta L_{q,g}$
refer to the quark and gluon angular momentum, while $\Delta G$ refers to the
gluon (hybrid) spin operator. As was shown in the previous section, the gluon
(hybrid) contribution to the ground state proton is small $(O(1\%))$ and we
shall neglect the last two terms in (\ref{72}). For the first two terms using
quark wave functions (\ref{68}) one obtains \be\left\lan
\chi_\uparrow|\Sigma_3| \chi_\uparrow\right\ran = \int^\infty_0
\left(G^2(r)-\frac13 F^2(r)\right)  dr= 1-\frac43 \eta,\label{73}\ee

\be\left\lan \chi_\uparrow|\Delta L_q| \chi_\uparrow\right\ran =
\frac23\int^\infty_0 F^2  dr,\label{74}\ee and hence \be  \left\lan
\chi_\uparrow|J_3| \chi_\uparrow\right\ran =\left\lan \chi_\uparrow|\frac12
\Sigma_3 + \Delta L_q| \chi_\uparrow
     \right\ran=
      \int[\frac12 \left(G^2-\frac13 F^2 \right) +\frac23  F^2]dr=\frac12 .\label{75}\ee

Hence one does not have proton spin problems in the rest frame, however the
quark orbital momentum is essential, indeed defining $\eta$ from (\ref{71}),
where $g_A=1,27$ one obtains that the first two terms contribute as follows \be
\frac12 \lan \Sigma_3\ran \simeq 0.38;  \lan \Delta L_q\ran \simeq
0.12.\label{76}\ee

We now turn to the polarized quark distributions (PQD) in the nucleon with the
spin  along $z$ direction, \be \Delta u(x) = (u_\uparrow (x) -
u_\downarrow(x)), ~~ \Delta d (x) = (d_\uparrow (x) - d_\downarrow (x))
etc.\label{77}\ee and the PQD  of the proton is \be g_1 (x) = \sum_i
\frac{e^2_i}{2} \Delta q_i (x) = \frac29 (\Delta u ( x) + \Delta \bar u (x)) +
\frac{1}{18} (\Delta d + \Delta \bar d + \Delta s + \Delta \bar
s).\label{78}\ee

One can connect PQD with  $g_A$,  namely \cite{9} \be g_A =  \int^1_0 [ \Delta
u + \Delta  \bar u - ( \Delta d + \Delta \bar d) ] dx.\label{79}\ee To
calculate $\Delta u (x, k_\bot)$ one can use Eq. (\ref{10}), where the square
brackets should be rewritten  for the chosen spin projection $\mu$ as \be [~~ ]
\to [~~]_{\mu, q}\equiv \sum_q \delta (x-x(q)) \delta (s_z(q) - \mu)
 \delta^{(2)} (k_\bot (q) - k_\bot),\label{80}\ee

 So that  for  $\Delta u$ one can write
$$ \Delta u (x, k_\bot) = \int \delta^{(2)} ( \sum^3_{i=1} \vek_{\bot i} )
 \prod^3_{i=1} d^2 k_{\bot i} dx_1 dx_2 dx_3 \delta (1-\sum^3_{i=1} x_i)
 \times$$$$\times \frac{M^3_0}{(2\pi)^3} \left|\Psi_N\left (\vek_{\bot 1},
  \vek_{\bot 2},\vek_{\bot
 3},
 M_0 \left( x_1 -\frac 13\right),  M_0 \left( x_2 -\frac 13\right),  M_0 \left( x_3 -\frac
 13\right)\right)\right|^2
$$
\be
 \frac12
  \left([~~]_{\frac12, u} - [~~]_{-\frac12, u}\right), \label{81}\ee
 and $\Psi_N$ for the proton is given by the sum  (\ref{67}) of the products of
 single-quark wave functions (\ref{68}).

 Both contributions $(\Sigma_3$ and $\Delta L_q$)  are boost and gauge invariant
 \cite{9} and using the PQD of (\ref{81}) and  our result of the previous
 section, that the Fock sequence $\{C^N\}$ is boost invariant, one can
 conclude, that the proton spin condition (\ref{73}) is satisfied also in the
 boosted system. One of the main conclusion of the previous section retains,
 that the Fock sequences of the ground state nucleon and the DIS Fock sequence
 refer to different objects and hence the ``proton spin problem'' with DIS
 data is actually the  ``excited baryon spin problem''.

To understand how the proton spin
 problem can be solved by the excited baryon states, we shall use again, as
in section 4,
 our baryon multihybrid model. In this case the Fock column state can be written as

\be \Psi_B = \sum^\infty_{N=1} C_N \Psi_{3q} + C_0  \Psi_{3q}, \label{82} \ee
 with the normalization condition (\ref{43}),
  and we shall as in section 4 neglect $C_0$ as compared to
  $$|C_N|^2 = 2.62 N^{-3/2}$$.

  To understand the spin structure of
   $\Psi_N$ we shall take into account that the spin-spin interaction
  is attractive for the opposite spin
   directions both for quarks and gluons, and therefore $N$ gluons form
  pairwise spin zero scalars, while an
   odd gluon on the string, connecting quark to string junction, can
  form preferrably the total quark-gluon
   spin $J=1/2$, so that the quark-gluon wave function can be written as

  \be \Psi_{qg}(1/2,1/2)
   = \sqrt{1/3} \chi_{1/2}^q \phi_0^g + \sqrt{2/3} \chi_{-1/2}^q \phi_1^g, \label{83} \ee
  and this combination yields $$<\Psi_{qg}(1/2,1/2)|\Sigma_3|\Psi_{qg}(1/2,1/2)> = -1/6. $$
  Hence the total sum over $N$ in (\ref{82}) can be split even N part,where gluons are paired
   and $<\Sigma_3> = 1/2$,
   and the odd $N$ part, yielding $-1/6$, so that the total answer is

  $$<\Psi_B|\Sigma_3|\Psi_B>
   = 1/2 \sum^\infty_{n=1} (2n)^{-3/2} -
    1/6 \sum^\infty_{n=0}
     (2 n + 1)^{-3/2} = $$ \be=(2^{-5/2} - 1/6(1- 2^{-3/2})) 2.62 = 0.183. \label{84} \ee

  It is remarkable that this result roughly agrees with the recent data from \cite{31}
   obtained for $Q^2 = 3 (GeV/c)^2$.
    Thus $\Sigma_3$ refers rather t the multihybrid state and not to the ground state proton.

  Our conclusion does not disprove or invalidate the enormous experimental and
 theoretical efforts, which have provided important information on DIS
 structure functions. The latter  can  be used in many proper places, where the
 intermediate high excited baryon states appear.

 \section{Conclusions and prospectives}

 We have presented in the paper the new formalism for the calculation of the
 boosted valence wave functions of mesons and baryons. It was shown above, how
 one derives the valence parton distributions from these wave functions, which
 can be called the proper valence parton distributions. We show, that the Fock sequence of a
 hadron is boost invariant,  and consequently contains the same dominant
 components, as in the rest frame.

 The Fock sequence of the DIS affected hadron is much more rich in higher
 components, the latter are given
  e.g. by the DGLAP
   formalism with a proposed initial state, or by the Fock space Hamiltonian.

 Another result of our approach is the nonperturbative character of the higher Fock
 components, which can be used as an initial step in the DGLAP  or BFKL evolution.

 The possible importance of the nonperturbative input can be seen in many examples
 of inconsistencies of the purely perturbative parton model, such as the high
  $p_t$ hadron reactions, Drell-Yan processes, breakdown of factorization
 theorems etc., see \cite{9*,32} for discussions and references.

 As a good check of our formalism we have chosen the proton spin problem, which
 was not resolved in the standard parton approach, using the DIS parton
 distributions. We have shown that this problem is solved in the proton c.m.
 system, where the admixture of antiquarks and gluons is negligible, and then
 have used the boost invariance of our parton distributions to formulate the
 same solution in an arbitrary system.

 To compare with the polarised DIS data we have used the multihybrid model to demonstrate the decreasing
 of $\Sigma_3$ due to gluon admixture.
 In fact,  the present paper together with the preceding one \cite{8}, is the
 first step in an attempt to construct  the new formalism of nonperturbative
 QCD at high momenta and  energies, especially in the highly boosted systems. As
 it is, we suggest the way, where the treatment of the  boost is extremely
 simple, so that one can directly reformulate the results obtained in the rest
 frame. This work for the PDF's of different baryons is  now in  the process \cite{33}.
  The next step: the interaction of two complexes boosted with respect to
 each other is a much more complicated problem, the first conclusions on the
 behavior of the decay amplitudes and formfactors were given in \cite{8}.

This work was supported by the RFBR grant 1402-00395. The author is grateful to
M.A.Trusov for the help in preparing of Fig.~1 and to  K.G.Boreskov, B.L.Ioffe,
O.V.Kancheli and A.G.Oganesian for useful discussions.

\end{document}